\documentclass[12pt]{article}

\usepackage{amssymb,epsfig,cite,amsmath}
\usepackage{tikz, graphicx}
\usepackage{xcolor}

\newcommand{\be}{\begin{equation}}
\newcommand{\ee}{\end{equation}}
\newcommand{\bea}{\begin{eqnarray}}
\newcommand{\eea}{\end{eqnarray}}

\newcommand{\vep}{\varepsilon}

\newcommand{\N}{\mathbb{N}}
\newcommand{\R}{\mathbb{R}}
\newcommand{\C}{\mathbb{C}}
\newcommand{\nn}{\nonumber}

\newcommand{\corresp}{\stackrel{\wedge}{=}}
\newcommand{\qeq}{\mathop{\stackrel{?}{=}}}

\newcommand{\zmone}{\stackrel{z \to -1}{\longrightarrow}}

\makeatletter
\@addtoreset{equation}{section}
\makeatother

\usepackage{cite}

\begin{document}
\vspace*{-10mm}
\begin{center}
  
{\large{\bf From Ramanujan to renormalization: the art of doing}}

\vspace*{4mm}  

{\large{\bf away with divergences and arriving at physical results}}

\vspace*{8mm}

Wolfgang Bietenholz \\
Instituto de Ciencias Nucleares \\
Universidad Nacional Aut\'{o}noma de M\'{e}xico \\
A.P.\ 70-543, C.P.\ 04510 Ciudad de M\'{e}xico, Mexico
\vspace*{2mm}

\end{center}

\noindent
{\em A century ago, Srinivasa Ramanujan --- the great self-taught
  Indian genius of mathematics --- died, shortly after returning
  from Cambridge, UK, where he had collaborated with Godfrey Hardy.
  Ramanujan contributed numerous outstanding results to different
  branches of mathematics, like analysis and number theory,
  with a focus on special functions and
  series. Here we refer to apparently weird values
  which he assigned to two simple divergent series, $\sum_{n \geq 1} n$
  and $\sum_{n \geq 1} n^{3}$. These values are sensible, however,
  as analytic continuations, which correspond to Riemann's
  $\zeta$-function. Moreover, they have applications in physics:
  we discuss the vacuum energy of the photon field, from which one
  can derive the Casimir force, which has been experimentally measured.
  We further discuss its interpretation, which remains controversial.
  This is a simple way to illustrate the concept of renormalization,
  which is vital in quantum field theory.}
\vspace*{-2mm}\\


\section{Ramanujan's letter from 1913}

101 years ago, Srinivasa Ramanujan\footnote{Phonetically,
  his last name could be written in Spanish as ``Ram\'{a}nuchan''.}
(1887-1920) passed away in Madras, at that time part of the British
Empire (since 1996, this state capital in
South-East India is named Chennai). He was one of the greatest geniuses
in the history of mathematics. One way to measure the impact of
  his work is through the amount of mathematical terms that bear his
  name: the
mathematical online encyclopedia {\it Wolfram Mathworld} \cite{mathworld}
documents 27 terms named after Ramanujan, and his name
appears in a total of 205 items; in both respects, he is among
the leading mathematicians of all times.\footnote{To be explicit,
  if we rank mathematicians by the number of mathematical items
  named after them, Ramanujan is at position 6, following Euler (71),
  Gauss (48), Hilbert (33), Fermat (32) and Riemann (31), and followed
  by Cauchy (26), Dirichlet, Jacobi, Weierstra\ss \ (23 each),
  Euclid (22) and Poincar\'{e} (21). Regarding the number of
  mentionings in a {\it Mathworld} entry, Ramanujan is at position
  18.\label{fn:ranking}}
This is particularly amazing since Ramanujan started to elaborate
stunning equations with hardly any mathematical
education,\footnote{Ramanujan only obtained from a friend a library copy
  of a book by George Carr \cite{Carr}, which he studied intensively.
  It is a collection of formulae and theorems, with little explanation,
  written as an overview for students who are preparing for
  exams.\label{fn:Carr}}
and he died at the age of only 32 (younger than Mozart, for example).

\begin{figure}[h!]
\begin{center}
 \includegraphics[angle=0,width=.4\linewidth]{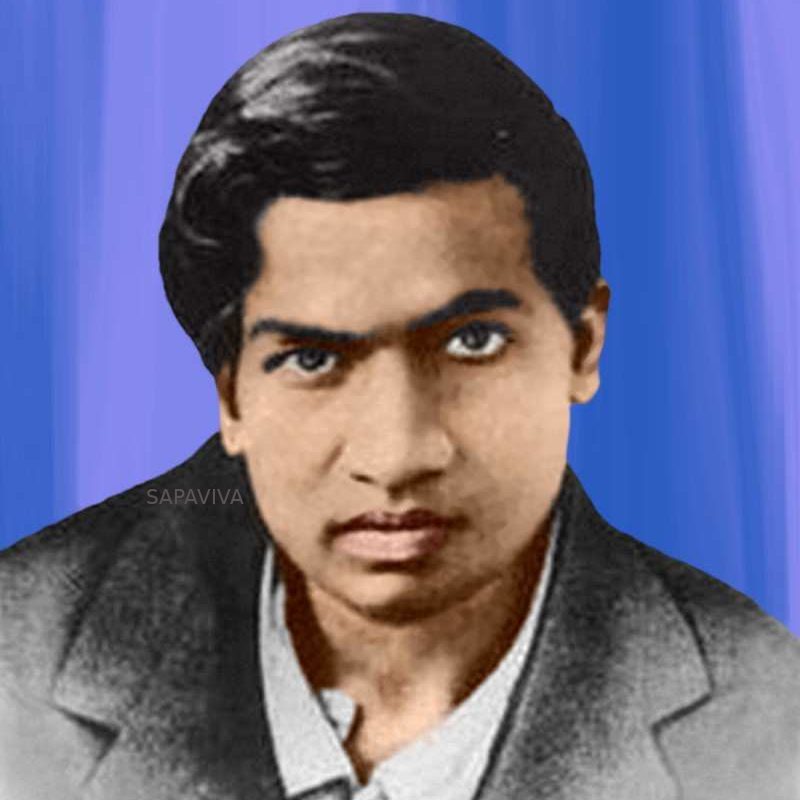}
\caption{\it Srinivasa Ramanujan (1887--1920)}
\end{center}
\vspace*{-3mm}
\end{figure}

Ramanujan was born in 1887 in a town called Erode, but at the age
of 2 his mother took him to the city of Madras, some 400~km away. 
In the early 20th century he lived in extreme poverty,
at the edge of starvation, but he discovered a multitude of important
mathematical formulae, based on his incredible intuition
--- I tend to interpret it as a kind of ``pattern recognition''
(although it was not an automated process).

In 1912, he began to send letters to British
mathematicians, trying to attract attention to his
discoveries; for a while without success.
In January 1913, he finally wrote to Godfrey Hardy, a brilliant young
mathematician at Trinity College of Cambridge University,
who --- together with his long-term collaborator John Littlewood
--- turned out to be the most influential British mathematician
of the first half of the 20th century. They are credited for
boosting British mathematics to the top level again, after it had
stayed behind the achievements in France and Germany during the 19th
century. In particular, Hardy insisted on mathematical rigor,
which was in total contrast to Ramanujan's intuitive
style \cite{Rambio}.

Unlike his colleagues, Hardy became aware of the enormous value
of Ramanujan's results, although part of it had been known before,
and some formulae were wrong --- but the rest was
groundbreaking \cite{Hardy20}. Having received two letters with 120
remarkable equations, Hardy invited
Ramanujan to Cambridge, which he accepted after some hesitation,
and where he stayed from 1914 to 1919, {\it i.e.}\ mostly
  during World War I. It was not easy for him
to get used to the climate, lifestyle and food.\footnote{Being a
  devout Hindu, Ramanujan was a strict vegetarian, which was highly
  unusual in England at that time.} Moreover, he suffered 
from serious health problems; they had antecedents in his earlier
life in India, and they lead to his decease one year after
his return to Madras.

Despite appreciating his brilliance, Hardy
urged him to take lectures (for instance, Ramanujan hardly knew
anything about complex analysis), and in particular he insisted
on {\em proofs,} not just conjectures. That was not
easily compatible with Ramanujan's mentality, but he published
32 high-impact papers during his 5 years in Cambridge,
7 of them together with Hardy \cite{Rpapers}. In 1918 Ramanujan
was elected as a Fellow of the Royal Society, as one of the youngest
members in its history, and 
half a year later he also became a Fellow of the Cambridge Trinity College.

\begin{figure}[h!]
\vspace*{1mm}
\begin{center}
  \includegraphics[angle=0,width=.4\linewidth]{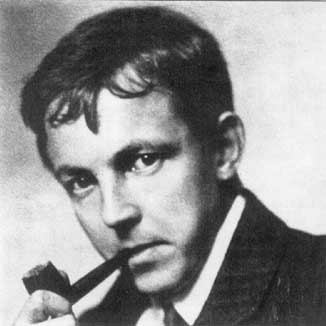}
\caption{\it Godfrey Hardy (1877--1947)}
\end{center}
\vspace*{-4mm}
\end{figure}

His investigations involved subjects,
which had been considered intractable before, in particular
a miraculous approximate formula for {\em partitioning,} which
(surprisingly) involves the number $\pi$ \cite{partitioning}.
Ramanujan traced this number in all kind of contexts; best known
is a series that he postulated in Ref.\ \cite{RamPi} (along with
a variety of other $\pi$-approximation formulae),
\be
\frac{1}{\pi} = \frac{\sqrt{8}}{99^{2}} \ \sum_{n \geq 0} \,
  \frac{(4n)!}{(4^{n} n!)^{4}} \
  \frac{1103 + 26390n}{99^{4n}} \ .
\ee
It converges exponentially (despite the factor $(4n)!$ in the numerator),
thus it provides one of the fastest algorithms to compute $\pi$.
If we truncate at $n_{\rm max} = 0,\, 1,\, 2$, we obtain the
corresponding approximation $\pi_{n_{\rm max}}$, which differs from the
exact value of $\pi$ as
\be
| \pi - \pi_{0}| \simeq 7.6 \cdot 10^{-8} \ , \quad
| \pi - \pi_{1}| \simeq 6.4 \cdot 10^{-16} \ , \quad
| \pi - \pi_{2}| \simeq 5.7 \cdot 10^{-24} \ .
\ee
{\em How} Ramanujan arrived at such formulae is hard to know:
Hardy later described it as a ``process of mingled argument,
intuition, and induction, of which he was entirely unable to give
any coherent account'' \cite{HardyZitat}.

Here we are going to address a relatively simple subject,
which Ramanujan mentioned in his first letter
to Hardy \cite{Rletter}, and which he had written down before in
Chapter VI of his Second Notebook \cite{Notebook}.
This letter contains two apparently weird formulae for divergent series,
\bea
\sum_{n \geq 1}^{\rm (R)} n = 1 + 2 +3 +4 + 5 + \dots = - \frac{1}{12} \ ,
\label{sumlin} \\
\sum_{n \geq 1}^{\rm (R)} n^3 = 1 + 8 + 27 + 64 + 125 + \dots = \frac{1}{120} \ ,
\label{sumcub}
\eea
where the sums run from $n=1 \dots \infty$, and the superscript
(R) indicates ``Ramanujan summation'' \cite{Candelpergher}.
These strange relations have fascinated generations of people;
for instance a discussion of eq.\ (\ref{sumlin}) in YouTube
\cite{YouTube}, dated 2016, has over 2.4 million views, and
over 5000 quite controversial comments.

Of course, it is provocative to write these relations as straight
equations, as Ramanujan did (without any superscript), but it fulfills
the purpose of attracting attention and causing debate.
Still, in the following we are
going to replace the symbol $=$ by $\corresp \,$, meaning
``corresponds to'' or ``is associated with''. In this sense,
we are going to show that the fractional values on the right-hand side
do have a meaning, not only as a mathematical peculiarity, but
they can even be used to derive {\em physical} results.

Unlike other addressees of Ramanujan's letters, Hardy recognized the
values of the {\em Riemann $\zeta$-function}, or {\em $p$-series.}
For ${\rm Re}\,z >1$ it is defined as
\be  \label{zetasum}
\zeta (z) = \sum_{n \geq 1} \frac{1}{n^{z}} = \frac{1}{1^{z}} +
\frac{1}{2^{z}} + \frac{1}{3^{z}} + \frac{1}{4^{z}} + \dots \ .
\ee
In 1739 Leonhard Euler had computed explicit expressions for
$\zeta (2n)$, $n \in \N_{+}$, and later he also conjectured a
$\zeta$-functional relation \cite{Eulerzeta}. More than a century later,
in 1859, Bernhard Riemann 
established the analytic continuation of the $\zeta$-function
to $\C - \{ 1 \}$ \cite{Riemannzeta}, see Appendix \ref{app:trust}.
In this sense, Hardy noticed that Ramanujan's results can be interpreted
as $\zeta(-1)$ and $\zeta(-3)$ (although these values are not explicitly
given in Ref.\ \cite{Riemannzeta}).

Riemann was a leading mathematician of the 19th century,
and of all times, cf.\ footnote \ref{fn:ranking}.
Like Ramanujan he lived his youth
in harsh poverty, until he was appointed
to a post in G\"{o}ttingen, on Carl Friedrich Gauss' recommendation.
Another analogy is that he soon suffered from health problems.
Hoping that a warmer climate might help against his
tuberculosis \cite{RiemannTuberculosis} (which was also among
Ramanujan's diseases \cite{Rambio}), he spent extended
periods in Italy, where he died in 1866, at the age of 39.

\begin{figure}[h!]
\begin{center}
\includegraphics[angle=0,width=.39\linewidth]{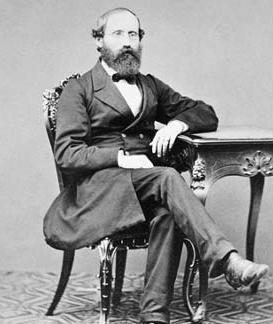}
\caption{\it Bernhard Riemann (1826--66)}
\end{center}
\vspace*{-4mm}  
\end{figure}

Differences from Ramanujan's life are that Riemann had access to
education at leading mathematical institutes, in G\"{o}ttingen and Berlin,
and that he published his results only after elaborating rigorous proofs.
His publications had an enormous impact,
but only Ref.\ \cite{Riemannzeta} deals with number theory.
There he discussed the density of prime numbers,\footnote{This
  is another field of common interest of these two geniuses:
  later Ramanujan proposed his own formula for the prime number density,
  which is, however, not as accurate as he had expected.}
and it was in this context that he postulated the analytic continuation
of the $\zeta$-function; the crucial functional
equation is displayed in eq.\ (\ref{zetafunctionaleq}).
In contrast to Ramanujan, Riemann was an expert on complex analysis.
Presumably he had hand-written notes with many more important
results, but after his sudden death his house-cleaner burned part
of these notes, until some mathematicians managed to stop
her \cite{Putzfrau}.

Ramanujan did not provide an actual derivation of formulae
(\ref{sumlin}) and (\ref{sumcub}), but in the first case
he assigned --- in Ref.\ \cite{Notebook} and also his first
letter to Hardy --- a value to another divergent series, as an
intermediate step to arrive at relation (\ref{sumlin}).
That series corresponds to a special case of {\em Dirichlet's
$\eta$-function,} or {\em alternating $\zeta$-function,}
\be  \label{etafun}
\eta (z) = \sum_{n \geq 1} \frac{(-1)^{n-1}}{n^{z}} = \frac{1}{1^{z}} -
\frac{1}{2^{z}} + \frac{1}{3^{z}} - \frac{1}{4^{z}} \dots \ ,
\ee
which converges for ${\rm Re} \, z > 0$. At ${\rm Re} \, z > 1$ we
obtain
\be \label{zetaeta}
\zeta (z) - \eta(z) = 2 \sum_{n \geq 1} \frac{1}{(2n)^{z}} =
2^{1-z} \zeta (z) \ , \quad \zeta (z) = \frac{1}{1-2^{1-z}} \, \eta (z) \ .
\ee
The latter defines $\zeta (z)$ in the domain
${\rm Re} \, z > 0 \wedge z\neq 1$.

In particular, Ramanujan wrote down its 
continuation to \cite{Notebook}
\be  \label{etaconst}
{\cal E} := \eta (-1) = \sum_{n \geq 1} (-1)^{n-1} n = 1 -2 + 3 - 4 \dots
\corresp \frac{1}{4} \ .
\ee
We are going to confirm this value, and follow his path to
relations (\ref{sumlin}) and (\ref{sumcub}), which we finally apply
to physics, in particular to the {\em Casimir effect.}

\section{Heuristic derivation of \ $\sum_{n\geq 1} n \corresp -1/12$}
\label{sec:anacon}

Series have both fascinated and confused mathematicians over and over
again, throughout history. The famous ``paradox'' by Zeno, which
describes a race between Achilles and a tortoise
(and further ``paradoxes'' of a similar style)
caused a deep crisis in the mathematics of Ancient Greece
(see {\it e.g.}\ Ref.\ \cite{Struik}), because the concept of
convergent series --- in this case, a geometrical series --- had
not yet been understood.

Here we just take the familiar geometrical series as the point
of departure. For $|z|<1$ we trivially obtain
\be
G(z) = \sum_{n \geq 0} z^{n} =
1 + z + z^2 + z^3 \dots = 1 + z \, G(z) \ \ \Rightarrow \ \
G(z) = \frac{1}{1-z} \ . \
\ee
The series converges only for $|z|<1$, but the final function $G(z)$
is defined all over $\C - \{ 1 \}$. Moreover, the complex function
$G(z)$ is {\em holomorphic} (or {\em complex analytic}, {\it i.e.}\ complex
differentiable)\footnote{We recall that this is a powerful
  property, which guarantees that the function has derivatives of any
  order in its domain of holomorphy, and that it coincides
  with its power series. Moreover, since $G'(z)\neq 0$ it is also
  {\em conformal,} {\it i.e.}\ angle conserving:
    if we interpret the function $G(z)$ as a map $\C \to \C$,
    and consider two curves $\gamma_{1}(z)$,  $\gamma_{2}(z)$, which
    intersect in $z_0$ with a certain angle, then the maps of these
    curves intersect in $G(z_{0})$ with the same angle.\label{fn:conform}}
in $\C - \{ 1 \}$, and therefore {\em meromorphic} in $\C$,
which implies that its analytic continuation from
the disk $|z|<1$ to $\C -\{1\}$ is unique, cf.\ Appendix \ref{app:trust}.

This allows us to define {\em Grandi's series}
\be
{\cal G} = 1 - 1 + 1 - 1 + 1 - 1 \dots = \sum_{n \geq 0} (-1)^{n} 
\ee
by means of analytic continuation,
\be
{\cal G} \corresp G(z)|_{z=-1} = \frac{1}{2} \ .
\ee
We can readily extend this scheme to the
$\eta$-function in eqs.\ (\ref{etafun}), (\ref{etaconst}). To this end,
we first return to safe grounds, {\it i.e.}\ to $|z|<1$, where
\be
G(z)^{2} = 1 + 2z + 3 z^{2} + 4 z^{3} \dots
= \sum_{n \geq 1} n \, z^{n-1} = G'(z) =\frac{1}{(1-z)^{2}} \ .
\ee
The function $G(z)^2$ is holomorphic as well, again with a (unique)
analytic continuation to $\C - \{ 1 \}$. This implies in particular
\be  \label{eta14}
   {\cal E} = \sum_{n \geq 1} (-1)^{n-1} n
\corresp G(z)^2|_{z=-1} = {\cal G}^{2} = \frac{1}{4} \ ,
\ee
which coincides with Ramanujan's result (\ref{etaconst}).

However, this is not yet what we need in order to assign a value
to the notorious series, which we denote as
\be  \label{Rseriesdef}
{\cal R} := \sum_{n \geq 1} n = 1 + 2 +3 +4 + 5 + \dots \ .
\ee
It would correspond to $G(z)^2|_{z=1}$, but $z=1$ is just the
point where this function has its double pole. Following
Ramanujan's line of thought \cite{Notebook}, we proceed by
introducing another series
\be  \label{R1def}
R_{1}(z) = 1 - 2z + 3z^2 - 4z^3 + 5z^4 \dots = \sum_{n \geq 1} n (-z)^{n-1} \ ,
\ee
which also converges at $|z|<1$, and we formally obtain
${\cal R} \corresp R_{1}(-1)$.\footnote{The reason for the notation with
  an index 1 will become clear in Appendix \ref{app:analyticon}.}
Again we refer to the safe region,
{\it i.e.}\ to the disk $|z|<1$, where we take the difference
\be
G(z)^{2} - R_{1}(z)
= 4z (1 + 2z^2 + 3z^4 + \dots ) = 4z \sum_{n \geq 1} n z^{2(n-1)}\ .
\label{Rrelation}
\ee
This operation can only be justified inside the
convergence disk, but once it is carried out, taking the
limit $z \to -1$ on both sides leads to
\be  \label{Rm12}
^{\lim}_{z \to -1} \Big[ G(z)^2 - R_{1}(z) \Big] = \frac{1}{4} - {\cal R}
\corresp -4 {\cal R} \quad \Rightarrow
\quad {\cal R} \corresp - \frac{1}{12} \ .
\ee
Thus Ramanujan removed the divergence in a controlled manner,
which leaves an unambiguous
finite value, and Hardy noticed that this assignment corresponds
to ${\cal R} = \zeta (-1)$. In Appendix \ref{app:analyticon} we will
discuss what has been going on here.

Alternatively we could invoke eq.\ (\ref{zetaeta}) and consider the
divergent series ${\cal C} = 1+1+1+1+1 \dots $. The limit $z \to 0$
implies
\be  \label{Cseries}
     {\cal C} \corresp \zeta (0) \corresp -\eta (0) \corresp
     -{\cal G} = -\frac{1}{2} \ ,
\ee
a value which is also given in Ramanujan's Second Notebook \cite{Notebook}.
When we even insert $z=-1$ in eq.\ (\ref{zetaeta}), we arrive again at
\be \label{CRseries}
   {\cal R} \corresp -\tfrac{1}{3} {\cal E} \corresp
   -\frac{1}{12} = \zeta (-1) \ .
\ee

So far this may look like a mathematical playground, but in the
next section we are going to apply this result to a physical toy
model, where it leads to sensible results.
In Section \ref{sec:Casimir3d} we will proceed to a setting, which
refers to physical phenomenology; for that purpose we will need
relation (\ref{sumcub}), which corresponds to $\zeta (-3)$.

\section{Casimir effect on a line}
\label{sec:toymodel}

  In this section and beyond, we are going to deal with {\em quantum field
  theory.} General introductions can be found in a number of textbooks,
  such as Refs.\ \cite{qft} (and a popular science description is given
  in Ref.\ \cite{WB}), but in order to follow the derivations in
  Sections \ref{sec:toymodel} and \ref{sec:Casimir3d} only very little
  knowledge about it is required. Our notation implicitly refers to
  the functional integral formulation, where the fields are functions
  of the space and time variables, with values which can, for instance,
  be real numbers (then it is a {\em neutral scalar field}, as
  in this section), or 
  vectors (as in Section \ref{sec:Casimir3d}).\footnote{Alternatively,
    in the canonical formalism the fields are operator-valued.}
  In general, all field
  configurations --- {\it i.e.}\ all possible values in each space-time
  point --- are integrated to obtain expectation values of observables.

  Here, however, we are only concerned with the ground state contributions
  of free fields. For a neutral scalar field we can imagine an (infinite)
  set of coupled harmonic oscillators, one at each space point.
  A Fourier transform yields oscillators
  for all possible frequencies. {\it A priori} these frequencies
  are not restricted, so if we sum up their ground state contributions
  to the vacuum energy density, the result diverges.

  We are going to be confronted with these
  ultraviolet (UV) divergences: they require a {\em regularization,}
  {\it i.e.}\ a mathematical modification which makes such sums
  (or integrals) finite, enabling calculations. In the end 
  we want to remove the regularization, hence we aim at a
  cancelation of the UV divergences.
  This can often be achieved by subtracting divergent terms, so-called
  {\em counterterms}, which correspond to some limit; without taking
  that limit, a finite quantity remains. This procedure is known
  as a {\em renormalization:} it should lead to finite results for
  the physical quantities, which do not depend on the regularization
  that has been chosen (if suitable conditions are fulfilled).

  The concepts, which we have sketched here in an abstract form,
  are going to be illuminated by the presentation of simple examples.\\
  
To this end, we address an effect,
which was theoretically predicted by the Dutch
physicists Hendrik Casimir and Dirk Polder in 1947/8 \cite{CasimirPolder}.
In particular, we follow the perspective that Casimir adopted a little
later \cite{Casimir48}, inspired by a discussion with Niels Bohr.

\begin{figure}[h!]
\begin{center}
\includegraphics[angle=0,width=.3\linewidth]{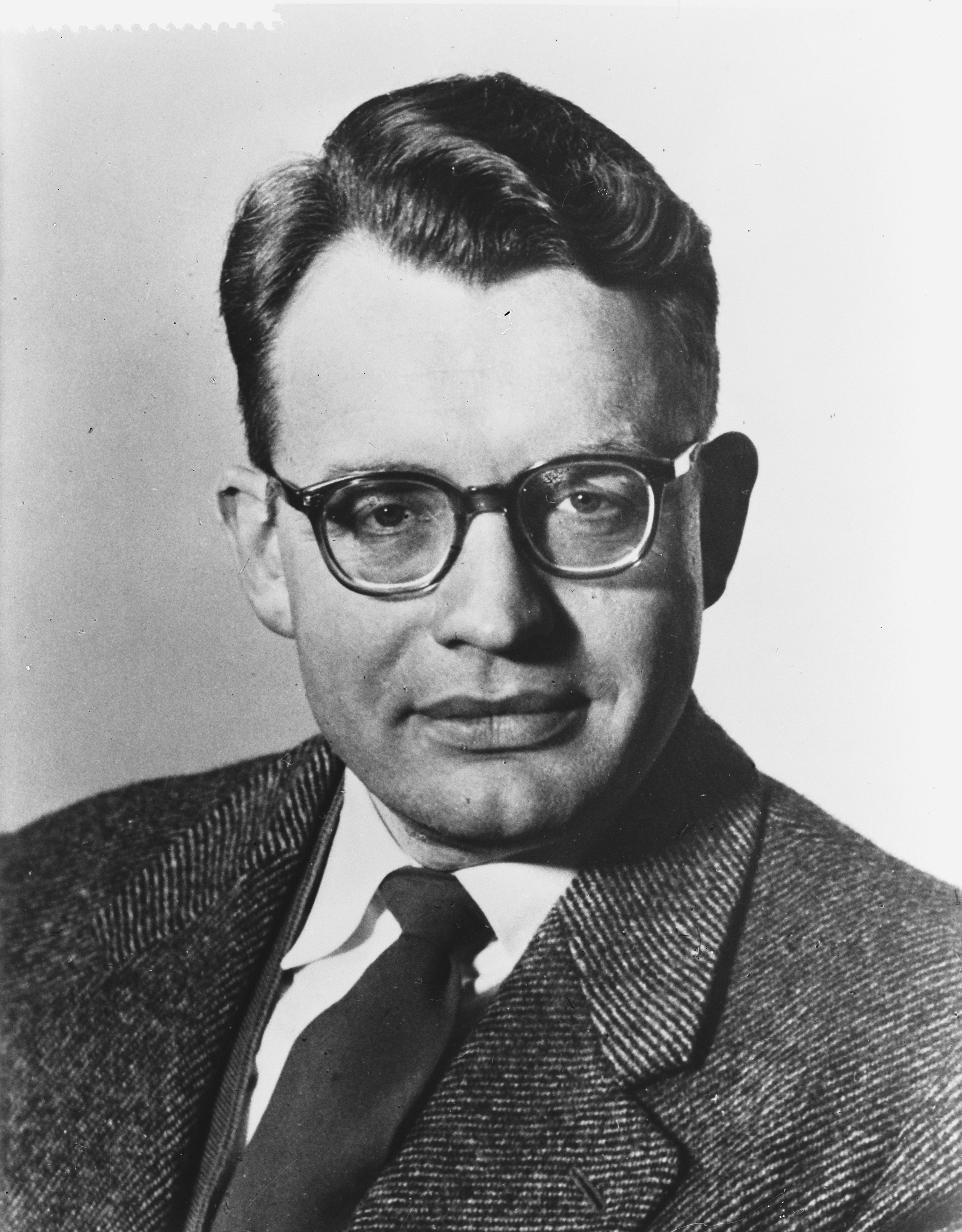}
\caption{\it Hendrik Casimir (1909--2000) }
\end{center}
\vspace*{-6mm}
\end{figure}

As a toy model, we consider a free, neutral, massless scalar field on a
line, $\phi(t,x) \in \R$, $x \in \R$. At the points $x=0$ and $x=d >0$ we
apply {\em Dirichlet boundaries,}
which force the field to vanish, {\it i.e.}
$\phi (t,0) = \phi(t,d)=0$. In this interval the field configurations
can be Fourier decomposed into standing waves with wavenumbers
$k_{n} = n \pi /d$, $n= 1,2,3 \dots$ (such that $\sin (k_{n}d)=0$).
In natural units ($\hbar = c=1$), they contribute $k_{n}/2$ to the
ground state energy.

For the {\em vacuum energy density} in this interval, we formally obtain
\be  \label{rho1d}
\rho_{d} = \frac{1}{2d} \sum_{n \geq 1} k_{n} = \frac{\pi}{2d^{2}}
\sum_{n \geq 1} n \ ,
\ee
where we encounter the divergent term ${\cal R}$, to which
Ramanujan assigned the value $-1/12$. We are now going to illustrate
--- in a physical framework --- why, and in which sense, this value is
indeed meaningful.

As usual in quantum field theory, we first need a {\em regularization}
(as we mentioned before), but it doesn't need to be fully specified.
We regularize $\rho_{d}$ by performing a substitution
\be  \label{fn}
n \to f(n) = n \, r(n/\Lambda d) \ , \quad {\rm with} \quad r(0) = 1 \ ,
\ee
where $f$ and $r$ are smooth functions on $\R_{0}^{+}$
(an infinite number of times
continuously differentiable), and $\Lambda$ is an energy cutoff.
If we remove it, $\Lambda \to \infty$, we recover the term before
regularization. At finite $\Lambda$ we require
\be  \label{fx}
^{\lim}_{x \to +\infty} \, f(x) = 0 \ , \quad
^{\lim}_{x \to +\infty} \, f^{(k)}(x) = 0 \ ,
\ee
where $f^{(k)}$ is any odd derivative $(k=1,\, 3,\, 5, \dots).$

A simple example of $f(n)$ in eq.\ (\ref{fn}) is the {\em heat kernel
  regularization}, where the function $r$ is exponential,
$f(n) = n \exp (-n /\Lambda d)$, which leads to a geometrical series,
\bea
\sum_{n \geq 1} n e^{-n /d\Lambda} &=& \Lambda^{2} d \
\frac{\partial}{\partial \Lambda} \sum_{n \geq 0} e^{-n /\Lambda d}
= \Lambda^{2} d \ \frac{\partial}{\partial \Lambda}
\frac{1}{1- \exp (-1/\Lambda d)} \nn \\
&=& (\Lambda d)^{2} - \frac{1}{12} +
     {\cal O}\Big( \frac{1}{\Lambda d} \Big) \ .
\label{heatkernel}
\eea
We already see ${\cal R} = -1/12$ popping up, but we want to proceed to
a broader perspective, which generalizes the regularization, and
which also clarifies the r\^{o}le of the UV divergent term, along
the lines of Ref.\ \cite{Casimir48}.

In an infinite interval, $d \to \infty$,
the formal term (\ref{rho1d}) for the energy density
turns into a momentum integral.
We expand the {\em difference} $\rho_{d} - \rho_{\infty}$,
at the regularized level, by means of the {\em Euler-Maclaurin formula,}
{\small{
\be  \label{EMformula}
\sum_{n=1}^{N} f(n) - \int_{0}^{N} f(x) \, dx =
\frac{f(N)-f(0)}{2} + \sum_{j \geq 1} \frac{B_{2j}}{(2j)!}
\Big( f^{(2j-1)}(N) - f^{(2j-1)}(0) \Big) \ ,
\ee  }}
\!\!\!where $N\in \N_{+}$. A finite number $N$ represents
  another component of the UV regularization: in this case, we sum
  over $k_{n}$ only up to $k_{n,{\rm max}} = N \pi /d$.

The powerful formula (\ref{EMformula}) was independently derived by Euler
and by Colin Maclaurin around 1735. It is very useful in field theory, in
particular when dealing with finite temperature or finite-size effects.
Since we assume the function $f$ to be smooth and to fulfill the
condition (\ref{fx}), this series converges both at finite $N$
and in the limit $N \to \infty$.\footnote{If we truncated
    this expansion at some odd integer $J/2$, such that we deal
    with $\sum_{j=1}^{J/2} \dots$, then there is a remainder
    term $R_{J}$ on the right-hand side. It can be estimated as
    $|R_{J}| \leq 2 \zeta(J) \int_{1}^{N}dx\ |f^{(J)}(x)| /(2\pi)^{J}$
    \cite{Apo}, hence the above assumptions imply
    $^{\lim}_{J \to +\infty} R_{J}=0$.}
The coefficients in the last term are the {\em Bernoulli
  numbers,}\footnote{Bernoulli numbers were a particular
  passion of Ramanujan, who had certainly read about them in Ref.\
  \cite{Carr}. His very first paper discussed their
  properties \cite{R1911}. For instance, he showed that
  the denominators of $B_{2}$, $B_{4}$, $B_{6}$, $B_{8}$ $\dots$
  (in lowest terms) all contain the prime factors 2 and 3 exactly once.}
which can be defined in a way related to eq.\ (\ref{heatkernel}),
\be
\frac{x}{1-e^{-x}} = \sum_{k \geq 0} B_{k} \frac{x^{k}}{k!} \ .
\ee
This yields $B_{0}=1$, $B_{1}=\frac{1}{2}$,
$B_{2}=\frac{1}{6}$, $B_{4}=-\frac{1}{30}$,
$B_{3}=B_{5}=B_{7}= \dots =0$,\footnote{This becomes obvious
    if we define $g(x) = x/(1-e^{-x})$ and compute $g(x) - g(-x) =x$.}
($B_{6}$, $B_{8} \dots$ do not vanish,
but we won't need them).

We insert in eq.\ (\ref{EMformula})
a function $f$ which fulfills the conditions
(\ref{fn}) and (\ref{fx}), and we take the UV limit in two steps:
  first we let $N\to \infty$; due to eq.\ (\ref{fx}) all contributions
  vanish in this limit.
As for the terms at $x=0$, we note that $f(0)=0$, $f'(0) = r(0) =1$,
$f^{(k)}(0) = {\cal O}((\Lambda d)^{1-k})$, $k = 2,\, 3, \dots$,
hence the second step of the UV limit, $\Lambda \to \infty$, leads to
\be  \label{rhorenorm}
\rho_{d} - \rho_{\infty} = \frac{\pi}{2d^{2}} \Big( -\frac{1}{2} B_{2} \Big)
= - \frac{\pi}{24 d^{2}} \ ,
\ee
where ${\cal R} = -B_{2}/2 = -1/12$ is crucial, in agreement with
eq.\ (\ref{heatkernel}).
We recover Ramanujan's assignment of a finite value to the divergent
series in eq.\ (\ref{rho1d}), {\it i.e.}\ {\em Ramanujan summation
corresponds to the subtraction of the infinite-volume limit
of the vacuum energy density.} This density diverges both in a finite
and in an infinite interval, but the {\em difference}, {\it i.e.}\
its finite-size effect, is finite and well-defined. The elimination,
or isolation, of a UV divergent term
($(\Lambda d)^{2}$ in eq.\ (\ref{heatkernel})), in order to deal with
finite differences, is the basic idea of {\em renormalization.}
In field-theoretic jargon, we have subtracted the {\em counterterm}
$\rho_{\infty}$, which cancels the divergence in the series (\ref{rho1d}).

Interestingly, this recipe matches exactly the relation
\be  \label{zetaminus1}
\zeta (-1) = \sum_{n\geq 1}^{\rm (R)} n = - \frac{1}{2} B_{2} = -\frac{1}{12} \ ,
\ee
applied to eq.\ (\ref{rho1d}).
We recall that the superscript (R) means {\em Ramanujan summation};
its general properties are defined and explored in Ref.\ \cite{Candelpergher}.
For the purpose of this article, it is sufficient to point out that
for a series of the form $\sum_{n\geq 1}^{\rm (R)} n^{k}$, $k \in \N$, the
Ramanujan summation coincides with the $\zeta$-function $\zeta (-k)$
(defined by analytic continuation), see Appendix \ref{app:trust}.
It further corresponds to the finite
term in the Euler-Maclaurin expansion of the difference
$^{\lim}_{N \to \infty} [\sum_{n=1}^{N} n^{k} - \int_{0}^{N} x^{k} \, dx]$,
which can be read off from eq.\ (\ref{EMformula}), and which
generalizes eq.\ (\ref{zetaminus1}) to
\be  \label{zetaminusm}
\zeta (-k) = \sum_{n\geq 1}^{\rm (R)} n^{k} = - \frac{B_{k+1}}{k+1}
\ , \quad k \in \N \ .
\ee

The question remains: beyond the satisfaction of deducing a finite
result, why are we interested in this difference?
What is its physical meaning? One is tempted to
reply: the change of the vacuum energy, as a function of $d$,
implies a force between the Dirichlet boundaries, and the
  counterterm can be subtracted since it does not depend on $d$,
  so it does not contribute to this force.
However, there is still a caveat, which is often ignored:
the boundaries could also affect the energy {\em outside} the
interval $[0,d\,]$. That could contribute to the force between
the boundaries, so we have to be careful.

\begin{figure}[h!]
\vspace*{-2mm}
\begin{center}
  \includegraphics[angle=0,width=.33\linewidth]{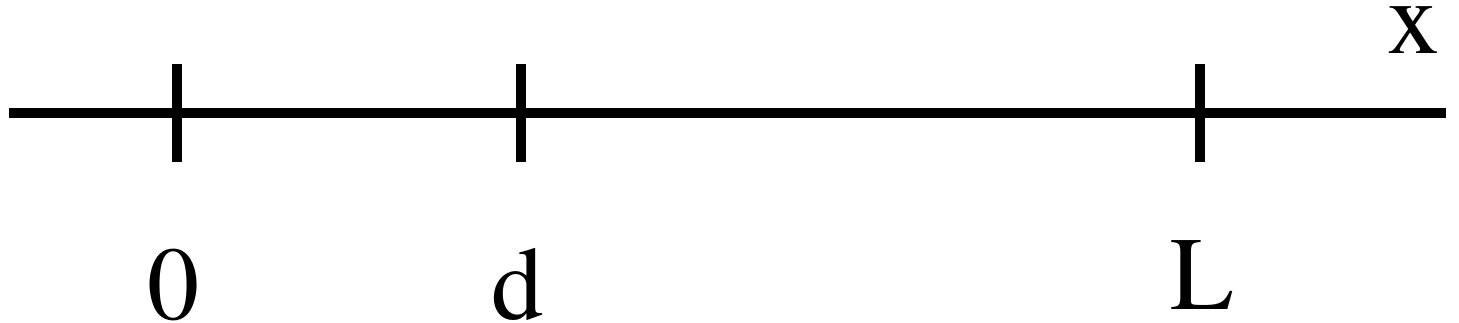}    
  \hspace*{1cm}
  \includegraphics[angle=0,width=.4\linewidth]{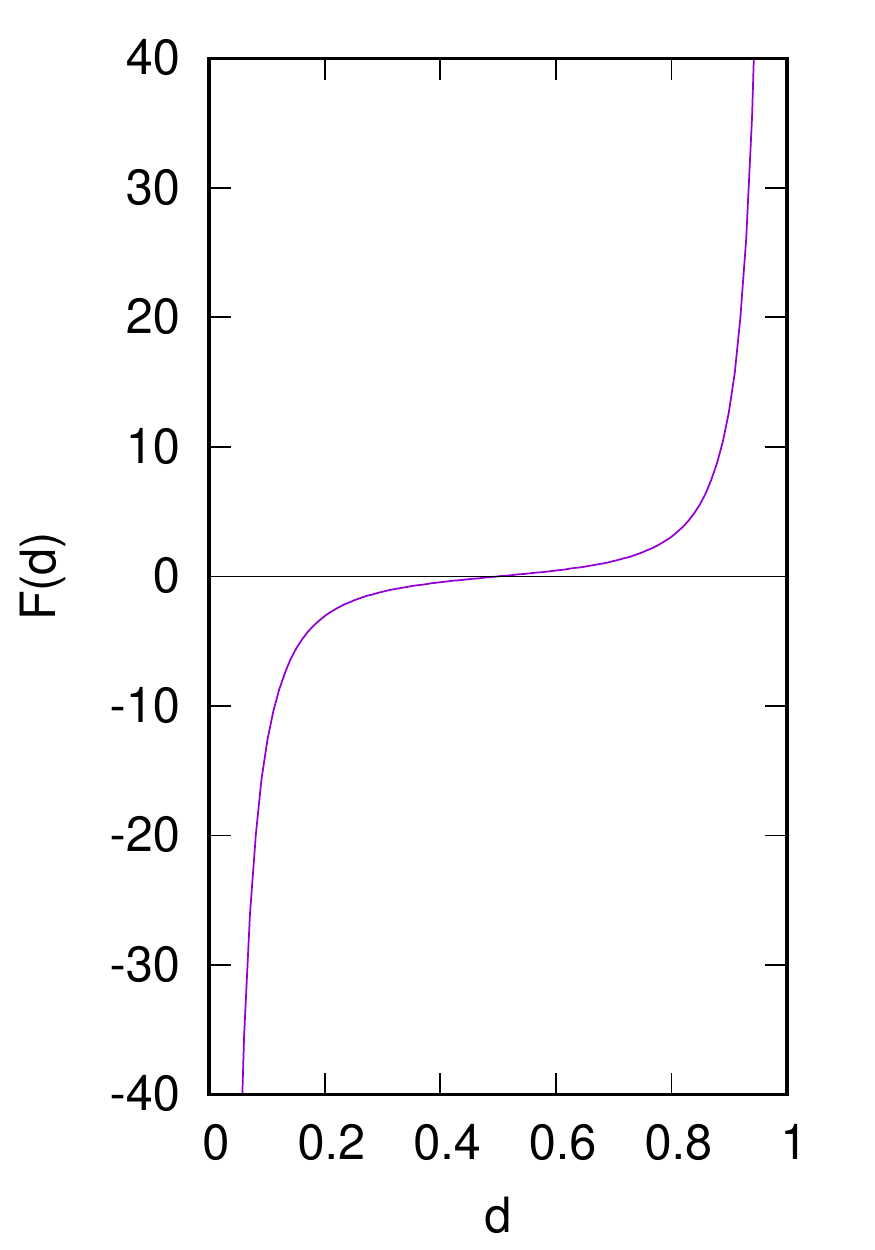}
\end{center}
\vspace*{-4mm}
\caption{\it Left: Setting for the Casimir effect on a line:
  we apply Dirichlet boundaries at the positions $0$, $d$ and $L$,
  with $0<d<L$. Right:
  The Casimir force $F(d)$, which acts on the
    ``piston'' at $d$ according to eq.\ (\ref{1dCasimirforce}),
  in units such that $L=1$.}
\vspace*{-1mm}
\label{Casimir1dfig}  
\end{figure}

A sound approach introduces {\em three} Dirichlet boundaries,
at the points $0$, $d$, $L$, with $0<d<L$, see Figure
\ref{Casimir1dfig} (left).
The idea is to keep the extreme boundaries at $0$ and $L$ fixed,
while the one at $d$ is a variable ``piston''.
In this way, the energy outside the interval $[0,L]$
  remains constant, while the energy inside this interval
  can be computed explicitly, so everything is under control.
From eq.\ (\ref{rhorenorm}) we obtain the total vacuum energy
\be
E(d) = -\frac{\pi}{24d} - \frac{\pi}{24(L-d)} + E_{\rm out}
= - \frac{\pi L}{24 d (L-d)} + E_{\rm out} \ .
\ee
The term $E_{\rm out}$, which represents the energy outside the
interval $[0,L\,]$, is divergent, but it does not
depend on $d$.
  Generally a force is obtained as the negative gradient of
  the potential energy. In our 1-dimensional case, this operator
  reduces to the negative derivative with respect to $d$ (the only
  variable involved). Therefore the term $E_{\rm out}$ is irrelevant
  for the force acting on the ``piston'' at $d$, which is obtained as
\be  \label{1dCasimirforce}
F(d) = - E'(d) = -\frac{\pi L}{24} \frac{L - 2d}{d^{2}(L-d)^{2}} \ ,
\ee
and depicted in Figure \ref{Casimir1dfig} (right).
It is odd with respect to the center $d=L/2$, and
{\em attractive} towards the nearer fixed boundary, at $0$ or $L$.
Hence $d=L/2$ is an unstable equilibrium position.
In the case $L \gg d$ we obtain a force, which is attractive
towards the boundary at $0$, $F(d) \simeq - \pi /(24 d^{2})$,
and which coincides with the 2-boundary picture of eq.\ (\ref{rhorenorm}).
Hence, in that picture, ignoring effects outside the interval $[0,d\,]$
is justified after all
(varying $d$ does not change the energy in the half-line with $x>d$).

To summarize this section, we have {\em renormalized} the system by
discarding an additive, infinite constant in the energy density,
the {\em counterterm} $\rho_{\infty}$,
which represents the infinite-volume limit.
In order to compute the remaining finite term ---
a finite-size effect in this case --- a regularization
  is needed. Then the Euler-Maclaurin formula can be applied, and by
  removing both UV cutoffs, $k_{n,{\rm max}} = Nd/\pi \to \infty$ and
  $\Lambda \to \infty$, we arrive at the finite result
  (\ref{rhorenorm}). It does not depend on the choice of the
  regularizing function $f$, as long as the conditions
  (\ref{fn}) and (\ref{fx}) are fulfilled. This leads to finite
values for the vacuum energy in the interval $[0,L]$,
and for the force $F(d)$ in eq.\ (\ref{1dCasimirforce}), which
acts on the ``piston''.

\section{The Casimir force in 3-dimensional space}
\label{sec:Casimir3d}

\begin{figure}[h!]
\begin{center}
  \includegraphics[angle=0,width=.18\linewidth]{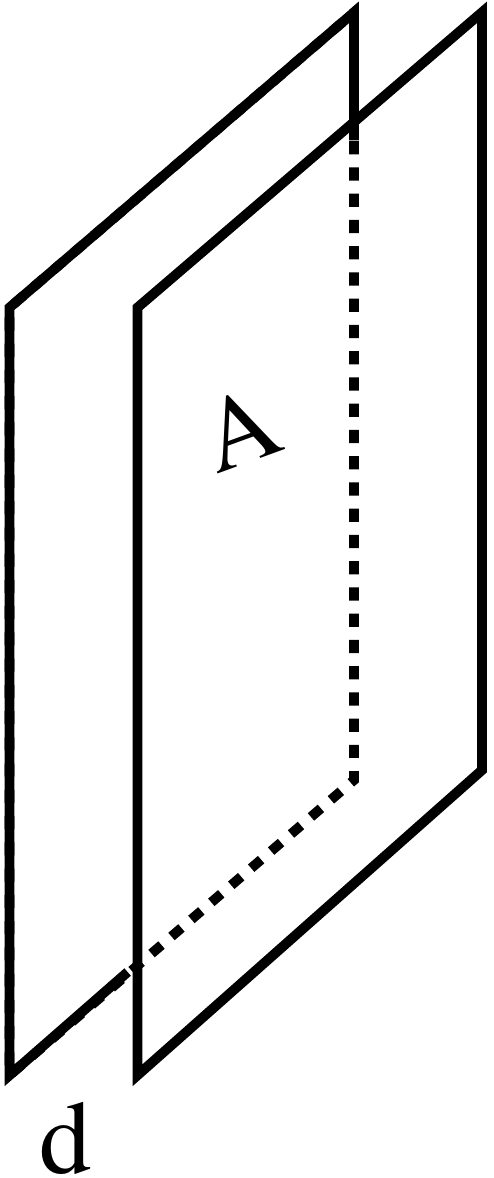}
\end{center}
\vspace*{-3mm}  
\caption{\it Setting for an experimental test of
    the Casimir effect: one measures the force 
    between two parallel, conducting plates of area $A$, separated
    by a short distance $d$.}
\label{plates}
\end{figure}

We proceed to a realistic situation, which deals with the vacuum
energy of the photon field in $(3+1)$-dimensional space-time. The
simplest setting is shown Figure \ref{plates}: it involves two
parallel, conducting plates,\footnote{In theory, we assume perfect
  conductivity, this is what it takes to implement exact Dirichlet
  boundaries. The experiments have been performed with well-conducting
  metal plates, which provide a good approximation,
  cf.\ Section \ref{sec:casimirforce}. The generalization with respect
  to the dielectric constant was theoretically studied by Evgeny
  Lifshitz \cite{Lifshitz56}.}
of the same rectangular shape and area $A$,
separated by a short distance $d$.\footnote{Throughout this article,
    we refer to the standard scenario with static Dirichlet boundaries. The
    two-fold generalization of the Casimir effect with dynamical Robin
    boundaries is discussed for scalar fields in
    Refs.\ \cite{dynbound}.}
  
We rely on our experience from the 1-dimensional toy model
to conjecture that it is sufficient to consider the energy $E(d)$
{\em between} the plates. This is appropriate when the area $A$
is large ($\sqrt{A} \gg d$). The photon momentum components
parallel to the plates --- which we denote as $k_{1}$, $k_{2}$ ---
are treated as continuous. Hence we perform a discrete sum, in analogy
to eq.\ (\ref{rho1d}), only over the vertical component $k_{3}$,
and the energy between the plates takes the form
\bea
&& \hspace*{-7mm} E(d) = A \, d \, \rho (d) =
\frac{1}{2} A \frac{1}{(2 \pi)^{2}} \int dk_{1} dk_{2} \
\sum_{n \geq 0} 2 \sqrt{k_{1}^{2} + k_{2}^{2} +
  \Big( \frac{\pi n}{d} \Big)^{2} } \nn \\
&& \hspace*{-7mm}  = \frac{A}{2\pi} \sum_{n \geq 0} \int_{0}^{\infty} dK \
K \sqrt{ K^{2} + \Big( \frac{\pi n}{d} \Big)^{2} }
= \frac{A}{6 \pi} \sum_{n \geq 0} \left.
\Big[ K^{2} + \Big( \frac{\pi n}{d} \Big)^{2} \Big]^{3/2}
\right|_{0}^{\infty} \ , \qquad
\label{Ed3d}
\eea
where we have inserted a factor $2$ for the two
photon polarization states.

Note that we have not regularized so far. If we do so and follow
the procedure of Section \ref{sec:toymodel}, we can renormalize by
subtracting the energy in the same volume but without plates,
$E_{\infty} = A\, d \, \rho_{\infty}$.\footnote{To obtain
  $E_{\infty}$ we start from the term in the upper line of
  eq.\ (\ref{Ed3d}), convert (in the large-$d$ limit) $\pi n/d$
  to the continuous momentum component $k_{3}$, and $\sum_{n \geq 0}$ 
  to $(d/\pi) \int_{0}^{\infty} dk_{3}$, which leads to
  $E_{\infty} = A\, d \, (2 \pi )^{-3} \int d^{3}k \, | \vec k |$.} 
In this difference, first the UV contribution due to $K \to \infty$
cancels (a physical interpretation is that infinitesimally
short wavelengths are not sensitive to the presence
of boundaries at a finite distance). Regarding $K=0$,
we apply the Euler-Maclaurin expansion (\ref{EMformula})
to \ $\sum_{n \geq 1} \dots - \int_{0}^{\infty} dk_{3} \dots$.
This corresponds to the Ramanujan summation over $n$, which we again
express as a $\zeta$-function,
\bea
\frac{E(d)}{A} & \corresp & - \frac{1}{6\pi} \frac{\pi^{3}}{d^{3}}
\zeta (-3) = -\frac{\pi^{2}}{720 d^{3}} \ , \nn \\
\frac{F(d)}{A} &=& \frac{-E'(d)}{A} = - \frac{\pi^{2}}{240 d^{4}} \ .
\label{3dCasimirforce}
\eea
We have used eq.\ (\ref{zetaminusm}) and inserted $B_{4}= -1/30$,
in agreement with eq.\ (\ref{sumcub}).
Although the sign of $B_{4}$ is opposite to $B_{2}$ (which we inserted
in eq.\ (\ref{rhorenorm})), we obtain again an {\em attractive} force
between the Dirichlet boundaries:
note that there is another sign flip due to the integral over $K$,
where the lower bound contributes.

The question arises of what magnitude this force takes for realistic
sizes $A$ and $d$, and if such a force can be measured.
The first conclusive experiment was achieved in 1997 by Steve
Lamoreaux, who succeeded in measuring the Casimir force to $5\,\%$ accuracy
\cite{Lamoreaux}. This was soon followed by Umar Mohideen and
Anushree Roy \cite{MohideenRoy}, who took into account the corrections
due to finite temperature, finite conductivity and the roughness
of the surfaces. In these experiments,
the geometrical structure was a plate and a sphere, because of the
difficulty in keeping two plates parallel to very high precision.

The first experiment to successfully measure the Casimir force between
parallel plates, as sketched in Figure \ref{plates}, was carried out
at the University of Padua, Italy, in 2002 \cite{Padua}, with
$15 \, \% $ precision. 
That experiment used rectangular silicon stripes, covered with a
chromium layer, of size $A = 1.9\, {\rm cm} \times 1.2 \, {\rm mm}$,
and their separation $d$ varied from
$0.5 \, \mu {\rm m}$ to $3 \, \mu {\rm m}$.
In order to compute the predicted force in Newton (N), we have to insert
a factor $\hbar c$ in eq.\ (\ref{3dCasimirforce}),\footnote{We recall
  that we have been using natural units. This factor shows that
  we are dealing with a relativistic quantum effect.}
which leads to
\be
F(d) \simeq -1.3 \cdot 10^{-7} \, {\rm N} \, \Big( \frac{\mu {\rm m}}{d}
\Big)^{4} \, \frac{A}{{\rm cm}^{2}} \ .
\ee
Hence the predicted force in this experiment varied between
$F \simeq -4.7 \cdot 10^{-7} ~{\rm N}$ and $-3.7 \cdot 10^{-10} ~{\rm N}$.
Forces of this range are in fact measurable:
for instance Ref.\ \cite{Lamoreaux} used a torsion
pendulum and laser interferometry, and Ref.\ \cite{MohideenRoy}
  employed an atomic force microscope. In the experiment
  reported in Ref.\ \cite{Padua}, one of the parallel plates
  was the face of a cantilever beam, free to oscillate.
  The variation of the force was observed by measuring shifts
  in its resonator frequency, employing a fiber-optic interferometer.

  The precision in recent experiments is around, or below, $1\,\%$.

\section{Concluding remarks}

We don't know what exactly Ramanujan had in mind
when he introduced his summation of divergent series, which we
now denote as Ramanujan summation, such as relations
(\ref{sumlin}) and (\ref{sumcub}).
In his letter to Hardy he only documented one intermediate step,
relation (\ref{eta14}). In his Second Notebook \cite{Notebook}
he additionally hinted at the continuation (\ref{Rm12}), and he
mentioned the difference between summation and
integration, which we also reviewed.
This is a valid argument, and apparently Ramanujan
reinvented an equivalent form of the Euler-Maclaurin formula
(he did not use that term \cite{Notebook}, nor does this
formula appear in Ref.\ \cite{Carr}).

A reason for the sparse documentation in Ramanujan's notebooks was ---
in addition to his intuitive way of thinking --- that he mostly
worked on a slate and only wrote down 
final results on paper, which was valuable (in particular for him,
who was living in poverty). It is also conceivable that he was
influenced by Carr's telegram style \cite{Carr}, cf.\ footnote
\ref{fn:Carr}. In any case, his approach matches the analytic
continuation of the $\zeta$-function, at least with respect to
negative integer arguments, see eq.\ (\ref{zetaminusm}).
In fact, he also rediscovered the analytically continued $\Gamma$-function
with values in $\C - \{0, -1, -2 \dots \}$. He highlighted this idea
in the introduction of his first letter to Hardy \cite{Rletter},
unaware that this had been known before; in particular, Riemann had
used it in Ref.\ \cite{Riemannzeta}.

These finite values for divergent series may look like a mathematical
game, which is rather disconnected from reality.
However, it is possible to establish consistent rules for the Ramanujan
summation of divergent series 
by carefully dealing with properties like linearity and translation
\cite{Candelpergher}.
Moreover, we reviewed their striking application to physics, where
they enable the prediction of a force, which has in fact been measured.
The {\em meaning} of the Casimir effect will be discussed
in the following two subsections.

\subsection{Is the electromagnetic vacuum energy density real?}
\label{sec:rhoreal}

Numerous authors infer from the experimental observation of the
Casimir force {\em the existence of the vacuum energy of the photon
field,} $\rho_{\rm vac}$, as predicted by Quantum Electrodynamics (QED),
{\it e.g.}\ Refs.\
\cite{Lamoreaux,MohideenRoy,Padua,WeinbergMilton,Carroll,SaSta}.
For instance, Ref.\ \cite{SaSta} states that ``the existence of zero-point
vacuum fluctuations has been spectacularly demonstrated by the
Casimir effect.''
It does, however, not affect usual experiments, which only depend
on energy {\em differences,} not on the additive constant $\rho_{\rm vac}$.
Still, such an energy density throughout the Universe, known
as {\em Dark Energy,} is indeed manifest since it affects the
expansion of the Universe.

It corresponds to the {\em Cosmological Constant} in General Relativity:
  in its absence --- which was generally assumed from the 1930s to
  the 1990s --- the expansion of the Universe would be decelerated.
  However, at the very end of the 20th century it was observed that
  the expansion is {\em accelerated}.\footnote{This was
    concluded from the distance and redshift of a set of type Ia
    supernovae \cite{Uexpand}.}
  This is best described by a positive Cosmological Constant, which
  corresponds to a Dark Energy density of about
  $\rho_{\rm DE} \approx (0.002~{\rm eV})^{4}$.

  Unfortunately this value is totally incompatible with the
  vacuum energy density $\rho_{\infty}$ that we discussed.  
  First, $\rho_{\infty}$ seems to diverge, as we saw, but one might
  impose an UV cutoff in the integral $\int d^{3}k \ |\vec k|$,
  most naturally at the Planck energy. This leads to
  a finite value $\rho_{\rm Planck}$, which is, however, {\em much} too
  large, $\rho_{\rm Planck} /\rho_{\rm DE} = {\cal O}(10^{121})$.
  
  People who still believe in {\em supersymmetry} could argue that in
  a perfectly supersymmetric world the Dark Energy vanishes (since
  bosons and fermions appear in pairs of the same mass, and the
  fermionic ground state energy is negative, with the same absolute
  value \cite{qft}). However, even if supersymmetry exists,
  it has to be badly broken in our low-energy world (otherwise particles
  like the ``selectron'' would have been observed), and the required
  extent of breaking still implies a Dark Energy density, which exceeds
  $\rho_{\rm DE}$ at least by a factor ${\cal O}(10^{60})$ \cite{Carroll}.
  Hence any evidence for the existence of the QED photon field vacuum
  energy $\rho_{\rm vac}$ would be 
  puzzling.

  Julian Schwinger {\it et al.}\ computed the Casimir force by means
  of a source field technique, without any need to refer
  to $\rho_{\rm vac}$ \cite{Schwinger}.
  Part of the literature concludes from that work that the Casimir
  experiments do not necessarily imply the reality of $\rho_{\rm vac}$,
  which could be welcome as a remedy against the
  disastrous discrepancy by 121 order of magnitude.

  Still, the question remains whether or not relativistic quantum
  physics could be formulated without $\rho_{\rm vac}$.   
  If $\rho_{\rm vac}$ exists, in the field-theoretic sense,
  one might wonder whether the frequency of a photon is affected
  when it passes through regions of different $\rho_{\rm vac}$,
  {\it e.g.}\ when it transversally passes through a Casimir
  cavity, similar to Bernoulli's Principle in fluid dynamics.
  Regarding its vertical motion, there is even a prediction
  that the speed of light could be affected \cite{Scharnhorst}.
  
  If we wanted to construct a cavity between two conducting plates
  with $\rho_{\rm DM} = |\rho_{d}| = \pi^{2}/ (720 d^{4})$, we would need
  a separation of $d \approx 0.3\, \mu{\rm m}$, which happens to be
close to the minimal separation in the Padua experiment.

\subsection{The nature of the Casimir force}
\label{sec:casimirforce}  

  One could question what kind of force this really is.
  It does not seem to appear in the famous list of four
  forces, which can be described by gauge fields, nor does it
  match further interactions in the Standard Model of particle
  physics (Yukawa couplings and the Higgs field self-coupling).
  However, being an effect of the {\em photon field,}
  this force must ultimately be electromagnetic, although this is
  not explicit in the above discussion.
  From this perspective, it can be best described as a
  {\em van der Waals force}\footnote{We refer to the
      van der Waals force in the narrow sense, also
      known as London--van der Waals force, {\it i.e.}\ an
      induced, attractive, collective multipole interaction
      between (electrically neutral) atoms or molecules \cite{DLP}.}
  between the metal plates. This is
  the picture that Casimir and Polder originally had in mind
  \cite{CasimirPolder}.

  So, do we have two equivalent descriptions? This seems puzzling
  again: in the van der Waals picture, the force depends on
  the value of the electromagnetic coupling constant
  $\alpha = e^{2}/4\pi \simeq 1/137$, which does not appear
  anywhere in the discussion based on the vacuum energy.

  This point was analyzed in depth by Robert Jaffe and
  collaborators, see Ref.\ \cite{RJaffe} for a summary.
  They conclude that the Casimir effect is a relativistic
  quantum force between electric charges and currents,
  {\it i.e.}\ a retarded van der Waals force, which does not
  require $\rho_{\rm vac}$. In Jaffe's own words, ``Casimir
    effects can be formulated and Casimir forces can be computed
    without reference to zero point energies.''
  Thus they contradict the paradigm
  in this field, but this issue remains controversial.

  Ref.\ \cite{RJaffe} obtains a Casimir force, which does depend
  on $\alpha$, such that $F(\alpha =0) = 0$, and
  $F(\alpha \simeq 1/137)$ is the physical strength.
  In this exceptional case, even the limit $\alpha \to \infty$
  leads to a finite result: $F(\alpha \to \infty)$
  just matches the force obtained from $\rho_{\rm vac}$,
  which is often close to $F(\alpha \simeq 1/137)$.
  For instance, for copper plates separated
  by $0.5 \ \mu {\rm m}$ (which is experimentally realistic), the
  consideration with $\rho_{\rm vac}$ is a good approximation if
  $\alpha \gg 10^{-5}$ \cite{RJaffe}, which is easily accomplished
  by the phenomenological value.

  A wide-spread objection against that point of view refers to
  examples, where the consideration based on $\rho_{\rm vac}$
  leads to a {\em repulsive} Casimir force \cite{Boyer},
  {\it e.g.}\ for specific parallelepipeds \cite{parallel}.
  That feature is not easily encompassed by van der Waals
  forces.\footnote{On the other hand, Ref.\ \cite{DLP} predicted a
    repulsive Casimir--van der Waals--type force,
    which agrees with an experiment
    with interacting materials immersed in a fluid \cite{Munday}.}
    For instance Lamoreaux \cite{Lamoreaux} writes:
    ``the Casimir and van der Waals forces are quite different;
    the van der Waals force is always attractive, whereas the sign of
    the Casimir force is geometry dependent.''
  Ref.\ \cite{HJKS} disagrees and assigns the repulsive result
  to the negligence of cutoff effects.
  In fact, an approach by Ricardo Cavalcanti \cite{Cavalcanti},
  which is manifestly free of any cutoff dependence, only obtains
  attractive Casimir forces.
  
  Jaffe and his collaborators insist that the physical Casimir force
  is always attractive \cite{RJaffe,HJKS},
  and therefore compatible with the van der Waals picture.
  If this alternative to the paradigm --- as expressed in Refs.\
  \cite{Lamoreaux,MohideenRoy,Padua,WeinbergMilton,Carroll,SaSta} ---
  is correct, then the approach that we reviewed is not the most precise
  one, but it is still in agreement with the experimental results.

Thus, returning to the title of Subsection \ref{sec:rhoreal},
doubts persist about the physical reality of $\rho_{\rm vac}$ as encoded
in QED. There is a consensus, however, that we do not know how to
  theoretically derive the Dark Matter density $\rho_{\rm DE}$,
    and that we do not understand the enormous discrepancy from
    the vacuum energy predicted by quantum field theory.

So far our discussion in this subsection focused on the Casimir
    effect due to QED, which was described in Section \ref{sec:Casimir3d}.
    In principle, such an effect also exists for other gauge fields, but
    only for QED it is simple and instructive, in particular because
    the photon field does not self-interact. This is also the only
    case where the Casimir force is experimentally confirmed.

For instance in {\em Quantum Chromodynamics} (QCD) --- the gauge theory
of the strong interaction --- this effect is much less transparent
because of the complicated self-interaction of the gluon field \cite{qft},
which occurs since the QCD gauge group SU(3) is
non-Abelian.\footnote{For interacting quantum field theories, renormalization
  involves more than subtracting a divergent term, which was sufficient
  in Section \ref{sec:Casimir3d} for the free electromagnetic field.
  In the interacting case, one assigns renormalized values to the fields
  and their couplings, which are in general energy-dependent.}
    At low energy
    its behavior is dominated by non-perturbative effects, which are
    hard to compute, and which induce an intricate vacuum structure.
    Studies in Euclidean space often focus on the r\^{o}le of
    instantons \cite{vBaal}.
    For a static quark--anti-quark pair (which is an idealization),
    a multipole expansion has been applied to estimate the Casimir
    force \cite{Bhanot}. Another study \cite{Canfora} deals with the
    (restricted) Gribov--Zwanziger action.

Furthermore, there are numerous attempts to
  theoretically investigate the {\em gravitational} Casimir force,
  although this is a quantum effect and we do not have
    any (fully satisfactory) theory of quantum gravity. Numerous
    papers refer to unusual gravitation theories; studies which are
    (roughly speaking) close to the framework of General Relativity
    include Refs.\ \cite{gravity}. The question whether an experimental
    demonstration of such an effect, with gravitational wave mirrors,
    would prove the existence of gravitons is discussed (and negatively
    answered) in Ref.\ \cite{Pinto}.

\subsection{Further physical applications of Ramanujan summation}
\label{sec:string}

There are further applications of Ramanujan summation in the
perturbative expansions of quantum field theory, which can be treated
by the {\em $\zeta$-function regularization} \cite{zetabook}. It is an
alternative to dimensional regularization, which is most popular
in perturbation theory. The $\zeta$-function regularization removes
from the beginning the UV divergent terms in the Laurent series by
inserting $\zeta (-k)$, thus preventing the necessity of
counterterms.\footnote{Its mathematical equivalence to the heat kernel
  regularization, see Section~\ref{sec:toymodel},
  was demonstrated by Hardy and Littlewood \cite{HardyLittlewood}.}
Stephen Hawking advocated its application in curved space-time
\cite{Hawking}.

Applications of the $\zeta$-function in {\em bosonic string theory}
are reviewed in Refs.\ \cite{string}.
Ref.\ \cite{Schwartz} summarizes a key point as follows:
a particle mass $m$ is obtained as
\be
m^{2} = \frac{1}{\sigma} \Big[ j + \frac{D-2}{2}
  \sum_{k \geq 1}^{\rm (R)} k \Big] \ ,
\ee
where $\sigma$ is the string tension, $D$ is the space-time
dimension (such that a worldsheet lives in $D-2$ dimensions), and
$j$ is the string excitation number
(here also the Planck scale is set to 1). The term with the sum
over the modes $k$ represents the ground state energy $E_{0}$, where
one applies Ramanujan summation, $E_{0}= -(D-2)/24 \sigma$.
The case $j=1$ describes spin-1 particles with only two polarization
states, which must therefore be massless. This condition yields
the space-time dimension $D=26$, where bosonic string theory is
formulated \cite{string}.
The deeper reason is the requirement to cancel the conformal anomaly.
A detailed pedagogical description is given in Ref.\ \cite{Zwiebach}.\\

\noindent
{\bf Acknowledgement: }
I would like to thank Kimball Milton for instructive comments.    
This work was supported by UNAM-DGAPA-PAPIIT, grant number IG100219.

\appendix

\section{The failure of partial sums}
\label{app:partialsums}

Many controversial discussions about relations like
(\ref{sumlin}) --- for instance numerous comments on Ref.\
\cite{YouTube} --- refer to partial sums of a few summands.
In the framework of divergent series, separating them is
conceptually wrong and leads to contradictions.
It is entertaining to look at some examples, to see what one
should beware of, {\it e.g.}
\be
{\cal R} = 1 + (2+3+4) + (5+6+7) + (8+9+10) + \dots \qeq 1 + 9{\cal R}
~ \to ~ {\cal R} \qeq -1/8 \ , \ 
\ee
which deviates from Ramanujan's value. One might even feel tempted
to pay attention to this alternative value, since it is consistent
with blocks of any odd number $u \geq 3$ of summands,
{\small{
\bea
&& \hspace*{-7mm} {\cal R} = 1 + 2 + \dots + \frac{u-1}{2} +
\Big( \frac{u+1}{2} + \dots + \frac{3u-1}{2} \Big) +
\Big( \frac{3u+1}{2} + \dots + \frac{5u-1}{2} \Big) \nn \\
&& \hspace*{-3mm} + \dots \qeq \frac{u^{2}-1}{8} + {\cal R} u^{2}
\quad \to \quad {\cal R} \qeq - \frac{1}{8} \ .
\label{oddblocks}
\eea  }}
\hspace*{-1.5mm}However, we can show that this approach is even
intrinsically inconsistent by choosing blocks of an even number $g$
summands (where the boundary terms are equally divided between the blocks),
\bea
    {\cal R} &=& 1 + 2 \dots ( \frac{g}{2}-1) + \frac{g}{4}
+ \Big( \frac{g}{4} + ( \frac{g}{2}+1) \dots + (\frac{3g}{2}-1) +
\frac{3g}{4} \Big) \nn \\
&+& \!\! \Big( \frac{3g}{4} + (\frac{3g}{2}+1) \dots \Big)
+ \dots \qeq \frac{g^2}{8} + {\cal R} g^{2} \quad \to \quad
{\cal R} \qeq - \frac{1}{8} \frac{g^{2}}{g^{2} -1} \ , \qquad \quad
\eea
which only coincides with the claim (\ref{oddblocks}) in the limit
$g \to \infty$.

Of course, the applicability of partial sums can be disproved more
easily, {\it e.g.}\ in 
Grandi's series or Dirichlet's $\eta$-function, if we write them as
\bea
{\cal G} &=& (1-1) + (1-1) + (1-1) + \dots
    = 1 + (-1+1) + (-1+1) + \dots \nn \\
{\cal E} &=& (1-2) + (3-4) + (5-6) + \dots = 1 + (-2+3) + (-4+5) + \dots \nn
\quad \eea
which seems to suggest the contradictory values ${\cal G} \qeq 0$ or
${\cal G} \qeq 1$,
and ${\cal E} \qeq \mp (1+1+1+1 \dots ) \doteq \mp {\cal C}$.
Again we encounter the series ${\cal C}$, which also appeared in Ramanujan's
Second Notebook \cite{Notebook}, as we anticipated in eq.\ (\ref{Cseries}).
We will come back to it in Appendix \ref{app:analyticon}.
In that case, a division into blocks of $n$ summands seems to
suggest ${\cal C} = (1+\dots +1) + (1+ \dots +1) + \dots \qeq n {\cal C}$,
${\cal C} \qeq 0$ or $\infty$ (while separating $k$ summands
${\cal C} = k +{\cal C}, \ {\cal C}=\infty$).

\section{Analytic continuation from the unit disk}
\label{app:analyticon}

We now follow the scheme of Section \ref{sec:anacon} by writing
the series under consideration in terms of a variable $z \in \C$,
such that they converge for $|z| <1$, and the divergent series of
interest corresponds to the limit $z \to -1$. Working in the convergence
region $|z| <1$ avoids, for instance, the confusion with partial sums.
First, we repeat the geometrical series and Dirichlet's $\eta$-function,
\bea
G(z) \!\!\!\!\!\! &=& \!\!\!\!\!\!
1 + z + z^{2} + z^{3} + \dots = \frac{1}{1-z}
\ \zmone \ {\cal G} = 1-1+1-1 \dots \corresp \frac{1}{2}
\ . \nn \\  
\eta (z)  \!\!\!\!\!\! &=&  \!\!\!\!\!\!
1 + 2z + 3z^{2} + 4z^{3} + \dots = G(z)^{2} = G'(z) =
\frac{1}{(1-z)^{2}} \nn \\
& \zmone & {\cal E} = 1-2+3-4 \dots \corresp \frac{1}{4} \ .
\eea
As long as this limit is finite, taking the analytic continuation is
trivial. This allows us to perform operations, which remain
valid in the limit $z \to -1$, like
\bea
\eta(z) + G(z) &=& \frac{1}{z} [\eta(z) -1] =
2 + 3z+4z^{2} + \dots \quad \zmone \ \frac{3}{4} \nn \\
\eta(z) - G(z) &=& z \eta(z) =
z + 2z^{2} +3z^{3} + \dots \quad \zmone \ - \frac{1}{4} \ .
\eea
In this framework, the limit $z\to -1$ is well controlled.

This is not obvious anymore when we deal with the pole of the
geometrical series at $z=1$, in particular when we refer to the series
\be
{\cal C} = 1+1+1+1+1 \dots
\ee
which worried us before in Section \ref{sec:anacon} and
Appendix \ref{app:partialsums}.
Here we consider three different regularizing functions (at $|z|<1$),
\bea
C_{1}(z) &=& 1 -z + z^{2} - z^{3} + z^{4} \dots = \frac{1}{1+z} \ , \nn \\
C_{2}(z) &=& 1 + z^{2} + z^{4} + z^{6} \dots = \frac{1}{1-z^{2}} \ , \nn \\
C_{3}(z) &=& -(z + z^{3} + z^{5} + \dots) = -\frac{z}{1-z^{2}} \ .
\eea
We can involve the functions $C_{i}(z)$ in a variety of relations, such
as $G(z) = (1+z)C_{2}(z) = 1 - (1+z) C_{3}(z)$. Of course they work both
in the form of series and of functions, in agreement with
$C_{i}(z\to -1) \to \infty$. We can also build linear combinations
of the functions $C_{i}(z)$, for instance
\be
G(z) = C_{1}(z) - 2 C_{3}(z) = \frac{1}{1-z} \ ,
\ee
where the singularity at $z=-1$ is {\em removed.}
When we now insert ${\cal C} = C_{1}(z\to -1) = C_{3}(z\to -1)$,
and treat it as a finite constant, we obtain the finite value, which
indeed matches $\zeta (0)$, as we saw in eq.\ (\ref{Cseries}),
and which Ramanujan had reported \cite{Notebook},
\be  \label{Cmh}
{\cal C} \corresp -{\cal G} \corresp -\tfrac{1}{2} = \zeta (0) \ .
\ee

It was (apparently) a step of this kind that Ramanujan
performed to compute the famous series (\ref{Rseriesdef}),
${\cal R} = 1+2+3+4+5 \dots \ .$ Here we consider the regularizations
\bea
R_{1}(z) &=& 1 - 2z + 3z^{2} - 4z^{3} + 5z^{4} \dots = C_{1}(z)^{2}
= - C_{1}'(z) = \frac{1}{(1+z)^{2}} \ , \nn \\
R_{2}(z) &=& 1 + 2z^{2} + 3z^{4} + 4z^{6} \dots = \eta(z^{2}) =
\frac{1}{(1-z^{2})^{2}} \ .
\eea
We introduced $R_{1}(z)$ before in eq.\ (\ref{R1def}), and
we used it, at the regularized level, in identity (\ref{Rrelation}),
which we can now write in the compact form
\be  \label{etaRR}
\eta (z) = R_{1}(z) + 4z R_{2}(z) \ .
\ee
Again the singularity at $z=-1$ cancels on the right-hand side:
note the an expansion in $\vep = z + 1$ leads to different Laurent
series for $R_{1}(\vep )=1/\vep^{2}$ and
$R_{2}(\vep) = (1 + \vep + 3\vep^{2}/4)/4 \vep^{2} + {\cal O}(\vep)$,
such that the right-hand side of eq.\ (\ref{etaRR}) takes the expected
form $1/4 + {\cal O}(\vep )$.

If we insert ${\cal R} = R_{1}(-1)=R_{2}(-1)$,
and treat ${\cal R}$ as a finite constant, we retrieve
Ramanujan's famous result
\be
\frac{1}{4} \corresp -3 {\cal R} \quad \Rightarrow \quad
{\cal R} \corresp - \frac{1}{12} = \zeta (-1) \ .
\ee

However, this procedure only works when the terms are arranged
such that the limit of interest ($z \to -1$ in our case) is
{\em regular,} otherwise this step is not controlled.
Consider for instance the identity
\be  \label{Cfail}
R_{1}(z) + C_{1}(z) = 2 - 3z + 4z^{2} - 5z^{3} \dots =
\frac{1}{z} [ 1 - R_{1}(z)] \ .
\ee
If we now insert ${\cal C} = C_{1}(-1)$ and
${\cal R}=R_{1}(-1) = -\tfrac{1}{z}R_{1}(z)|_{z=-1}$, we end up with
${\cal C} \qeq -1$, which contradicts eq.\ (\ref{Cmh}).

Another example, which refers to ${\cal R}$, is the identity
\be \label{Rfail}
R_{2}(z) = R_{1}(z) \eta (z) = \frac{1}{(1+z)^{2} (1-z)^{2}} \ .
\ee
Carelessly inserting ${\cal R}=R_{1}(-1)=R_{2}(-1)$ purports
${\cal R} \qeq \tfrac{1}{4}{\cal R}$, ${\cal R} \qeq 0$ or $\infty$.
The reason is that eqs.\ (\ref{Cfail}) and (\ref{Rfail}) are singular
at $z=-1$.

Ramanujan had either the right intuition to pick an appropriate
relation,  (\ref{Rrelation}) or (\ref{etaRR}),
where this step works, or it is based on
additional considerations on his slate, which are not documented.
It seems that he hardly knew any literature about the $\zeta$-function,
but he rediscovered correct values of its analytic continuation to
$\zeta (0)$, $\zeta (-1)$ and $\zeta (-3)$, which cannot be by accident.
In particular, he must have observed \cite{Notebook} the agreement
of the results that he obtained in this way with the finite term
in the series that we call Euler-Maclaurin expansion,
cf.\ Section \ref{sec:toymodel}.

\section{The Riemann $\zeta$-function}
\label{app:trust}

We have seen that a na\"{\i}ve ansatz for the continuation
can be plagued by subtleties when we hit a pole.
An unambiguous approach to evaluate series like ${\cal C}$
and ${\cal R}$, as well as the cubic series (\ref{sumcub}),
combines the terms such that --- in the limit
of interest --- the singularity is removed, as it is done
in the approaches of eqs.\ (\ref{Rrelation}), (\ref{etaRR}), and
of eqs.\ (\ref{Cseries}), (\ref{CRseries}),
or by subtracting the corresponding integral:
the result, given in eq.\ (\ref{zetaminusm}), coincides
with $\zeta (-k)$, $k \in \N_{0}$.

The underlying concept is {\em analytic continuation:}
if a complex function $f(z) \in \C$,
$z \in \C$, is holomorphic in some region, then its analytic
continuation is unique. Hence the existence of a complex
derivative is a powerful property: a plausibility argument is
that such a map $f(z)$, with $f'(z) \neq 0$, is angle-preserving,
cf.\ footnote \ref{fn:conform},
which constrains the analytic continuation to a single possibility.

Being a pioneer in this field, Riemann extended the $\zeta$-function
from the region with ${\rm Re}\, z > 1$ (where it is defined by
the convergent series (\ref{zetasum})) to $\C - \{ 1\}$
by means of relations \cite{Riemannzeta}, which can be condensed
into the functional equation
\be  \label{zetafunctionaleq}
\zeta (z) = \frac{(2 \pi )^{z}}{\pi} \sin ( \frac{\pi z}{2} ) \,
\Gamma (1-z) \, \zeta (1-z) \ ,
\ee
which is {\em valid all over} $\C$.
We read off $\zeta (-1) = (2 \pi^2)^{-1} (-1) \pi^{2} /6 = -1/12$,
where we inserted $\Gamma (n) = (n-1)!$, $n \in \N_{+}$, as well
as Euler's Basel formula $\sum_{n \geq 1} 1/n^{2} = \pi^{2}/6$.
Similarly we obtain $\zeta (-3) = 6 \zeta (4)/ (8 \pi^{4})
= 1/120$, by employing Euler's result $\zeta (4) = \pi^{4}/90$.
This confirms again the Ramanujan summations (\ref{sumlin})
and (\ref{sumcub}). We further see that $\zeta (-2n)=0$,
$\forall n \in \N$ (due to the sin-function), in agreement
with eq.\ (\ref{zetaminusm}).\footnote{These are the ``trivial
  zeros'' of $\zeta (z)$. According to the famous Riemann Conjecture,
  all other (``non-trivial'') zeros have ${\rm Re}\,z = 1/2$.}
Finally we observe a simple pole at $z=1$, with the residue
$^{\lim}_{z \to 1} (z-1) \zeta (z) =1$,
which is consistent with $\zeta (0) = -\tfrac{1}{2} = {\cal C}$.

The validity of a {\em series representation} of $\zeta (z)$, which converges
all over $\C - \{ 1 \}$, was demonstrated by Helmut Hasse \cite{Hasse}.
This formula uses eq.\ (\ref{zetaeta}) and a double sum for $\eta (z)$,
\be
\zeta (z) = \frac{1}{1-2^{1-z}} \sum_{n \geq 0} \frac{1}{2^{n+1}}
\sum_{k=0}^{n} (-1)^{k} \left( \begin{array}{c} n \\ k \end{array} \right)
\frac{1}{(k+1)^{z}} \ .
\ee
For $z=0$ the sum over $k$ corresponds to $(1-1)^{n} = \delta_{n,0}$,
and we confirm $\zeta (0) = -1/2$. Similarly, for $z=-1$ the
second sum yields $\delta_{n,0} - \delta_{n,1}$, and we obtain
once more $\zeta (-1) = -1/12$.

So could we have directly quoted these values and skipped the
previous consideration? There are two objections:
we would not capture the magic of Ramanujan's way of thinking,
and we would have missed the physical picture which leads to
the values of $\zeta (-n)$, $n \in \N$. This picture justifies
their application to the Casimir effect as a basic example
of renormalization.

\end{document}